# Reversible Programmable Logic Array (RPLA) using Fredkin & Feynman Gates for Industrial Electronics and Applications

Himanshu Thapliyal* and Hamid R. Arabnia**
*Center for VLSI and Embedded System Technologies,
International Institute of Information Technology, Hyderabad-500019, India
**The University of Georgia, Department of Computer Science,
415 Graduate Studies Research Center, Athens, Georgia 30602-7404, U.S.A
(thapliyalhimanshu@yahoo.com, hra@cs.uga.edu)

*Abstract*— In recent years, reversible logic has emerged as a promising computing paradigm having application in low power CMOS, quantum computing, nanotechnology, and optical computing. The classical set of gates such as AND, OR, and EXOR are not reversible. In this paper, the authors have proposed reversible programmable logic array (RPLA) architecture using reversible Fredkin and Feynman gates. The proposed RPLA has n inputs and m outputs and can realize m functions of n variables. In order to demonstrate the design of RPLA, a 3 input RPLA is designed which can perform any $2^8$ functions using the combination of 8 min terms ($2^3$). Furthermore, the application of the designed 3 input RPLA is shown by implementing the full adder and full subtractor functions through it.

## I. INTRODUCTION

This section provides an effective background of reversible logic with its definition and the motivation behind it.

### A. Definitions

Researchers like Landauer have shown that for irreversible logic computations, each bit of information lost generates kTln2 joules of heat energy, where k is Boltzmann's constant and T the absolute temperature at which computation is performed [1]. Bennett showed that kTln2 energy dissipation would not occur, if a computation is carried out in a reversible way [2], since the amount of energy dissipated in a system bears a direct relationship to the number of bits erased during computation. Furthermore, voltage-coded logic signals have energy of $Esig = \frac{1}{2}CV^2$, and this energy gets dissipated whenever switching occurs in conventional (irreversible) logic implemented in modern CMOS technology. It has been shown that reversible logic helps in saving this energy using charge recovery process [12]. Reversible circuits are those circuits that do not lose information. Reversible computation in a system can be performed only when the system comprises of reversible gates. These circuits can generate unique output vector from each input vector, and vice versa, that is, there is a one-to-one mapping between input and output vectors. Thus, an NXN reversible gate can be represented as

Iv=(I1,I2,I3,I4,……………………IN)
Ov=(O1,O2,O3,…………………ON).

Where Iv and Ov represent the input and output vectors respectively. Classical logic gates are irreversible since input vector states cannot be uniquely reconstructed from the output vector states. There are a number of existing reversible gates such as Fredkin gate[3,4,5], Toffoli Gate (TG) [3, 4] and TSG gate[6].

### B. Proposed Contribution and Motivation Behind the Work

The reversible logic operations do not erase (lose) information and dissipate very less heat. Thus, reversible logic is likely to be in demand in high speed power aware circuits. Reversible circuits are of high interest in low-power CMOS design [7], optical computing [8], quantum computing [9] and nanotechnology [10]. The most prominent application of reversible logic lies in quantum computers. A quantum computer can be viewed as a quantum network (or a family of quantum networks) composed of quantum logic gates; each gate performs an elementary unitary operation on one, two or more two–state quantum systems called qubits. Each qubit represents an elementary unit of information corresponding to the classical bit values 0 and 1. Any unitary operation is reversible, hence quantum networks effecting elementary arithmetic operations such as addition, multiplication and exponentiation cannot be directly deduced from their classical Boolean counterparts (classical logic gates such as AND or OR are clearly irreversible).Thus, Quantum Arithmetic must be built from reversible logic components [11].

Furthermore, programmable logic arrays (PLAs) have a number of medical and industrial applications, such as ultrasonic flaw detection. The reasons stem from the fact that PLAs are considerably faster than high end DSPs. They provide the cost effective solution to the exponentially increasing needs of industrial electronics. Thus, seeing the benefits of programmable logic array and reversible logic in industrial electronics and applications, the authors propose the reversible programmable logic array (RPLA) designed using reversible gates. In order to demonstrate the proposed architecture of RPLA, a 3 input RPLA which can perform any $2^8$ functions using the combination of 8 min terms ($2^3$) is also designed. The applications of the designed 3 input RPLA is shown by implementing 1-bit full adder and 1-bit subtractor functions through it. The basic goals in reversible logic are to minimize the number of reversible gates and garbage outputs. Thus, the proposed RPLA is designed in an optimal manner by appropriately choosing the reversible Fredkin[3,4,5] and Feynman gates[3].

## II. BASIC REVERSIBLE GATES

There are a number of existing reversible gates in literature. Fredkin[3,4,5] and Feynman gates[3] are used to construct the reversible PLA. A brief description of the gates is given below.

### A. Fredkin Gate

Fredkin gate [3,4,5], is a (3*3) conservative reversible gate originally introduced by Petri [4,5]. It is called 3*3 gate because it has three inputs and three outputs. The input triple $(x_1, x_2, x_3)$ associates with its output triple $(y_1, y_2, y_3)$ as follows:

$$y_1 = x_1$$
$$y_2 = (\neg x_1 \wedge x_2) \vee (x_1 \wedge x_3)$$
$$y_3 = (x_1 \wedge x_2) \vee (\neg x_1 \wedge x_3)$$

Figure 1.a and Figure 1.b show the implementation of the Fredkin gate as OR and AND functions respectively.

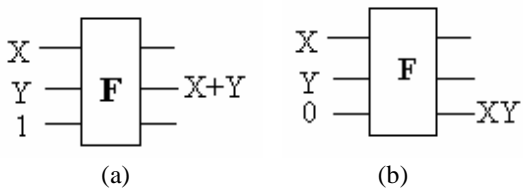

Figure 1. Fredkin Gate as (a) OR Function (b) AND Function

### B. Feynman Gate

Feynman gate [3] is a 2*2 one-through reversible gate shown in Figure 2. It is called 2*2 gate because it has 2 inputs and 2 outputs. One through gate means that one input variable is also the output. The input double (x1,x2) associates with its output double (y1,y2) as follows.

$$y1 = x1;$$
$$y2 = x1 \oplus x2;$$

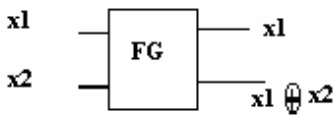

Figure 2. Feynman Gate

Figure 3a shows the implementation of Feynman gate for copying the input and Figure 3b shows the implementation of it for generating the complement of the input.

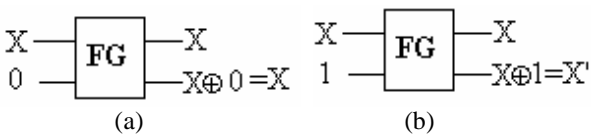

Figure 3. Feynman gate as (a) Copier (b) Complementer

## III. PROPOSED REVERSIBLE PROGRAMMABLE LOGIC ARRAY (RPLA)

In this paper, the authors have proposed the architecture of reversible PLA called RPLA. The architecture of the reversible PLA is shown in Figure 4. The RPLA consists of reversible AND array designed from reversible Fredkin gate and Feynman gate, in which the Feynman gates are used to avoid the problem of fan-out and for generating the complement of the inputs. The reversible AND array realizes the selected product terms of the input variables. The reversible OR array designed from reversible Fredkin gates is used to generate various possible functions of the product terms (outputs of reversible AND array).

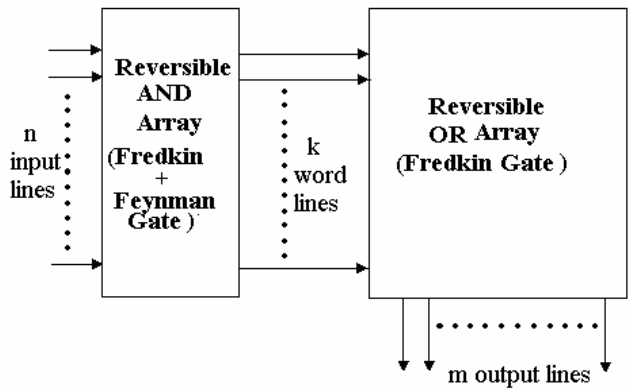

Figure 4. Proposed Reversible PLA (RPLA) Architecture

## IV. DESIGN OF 3 INPUT REVERSIBLE PLA (RPLA)

In order to demonstrate the actual design of the proposed reversible PLA (RPLA), a 3 input RPLA is shown in Figure 5. In Figure 5, the AND functions required to realize the AND array are implemented using reversible Fredkin gates. The application of Fredkin gate as an AND and OR gate has already been discussed in the previous section. In the AND array, the complement of the inputs are required and moreover fan-out is not allowed in reversible logic, thus Feynman gates are used to complement and replicate the signals when required. The designed 3 input reversible AND array will generate 8 product terms as outputs, which are combined using the reversible OR array designed using the Fredkin gate, to generate the required output functions.

### A. Applications of the designed 3 Input RPLA

The designed 3 input RPLA is used to implement the 1 bit full adder and 1-bit subtractor. The 1-bit full adder as shown in Figure 6 is implemented using the 3 input RPLA by generating the product terms in the full adder truth table through the AND array, and then appropriately combining the product terms through the reversible OR array to finally generate the required SUM and CARRY output functions. Similarly, the 1-bit subtractor as shown in Figure 7 is implemented to generate the Difference and Borrow output functions.

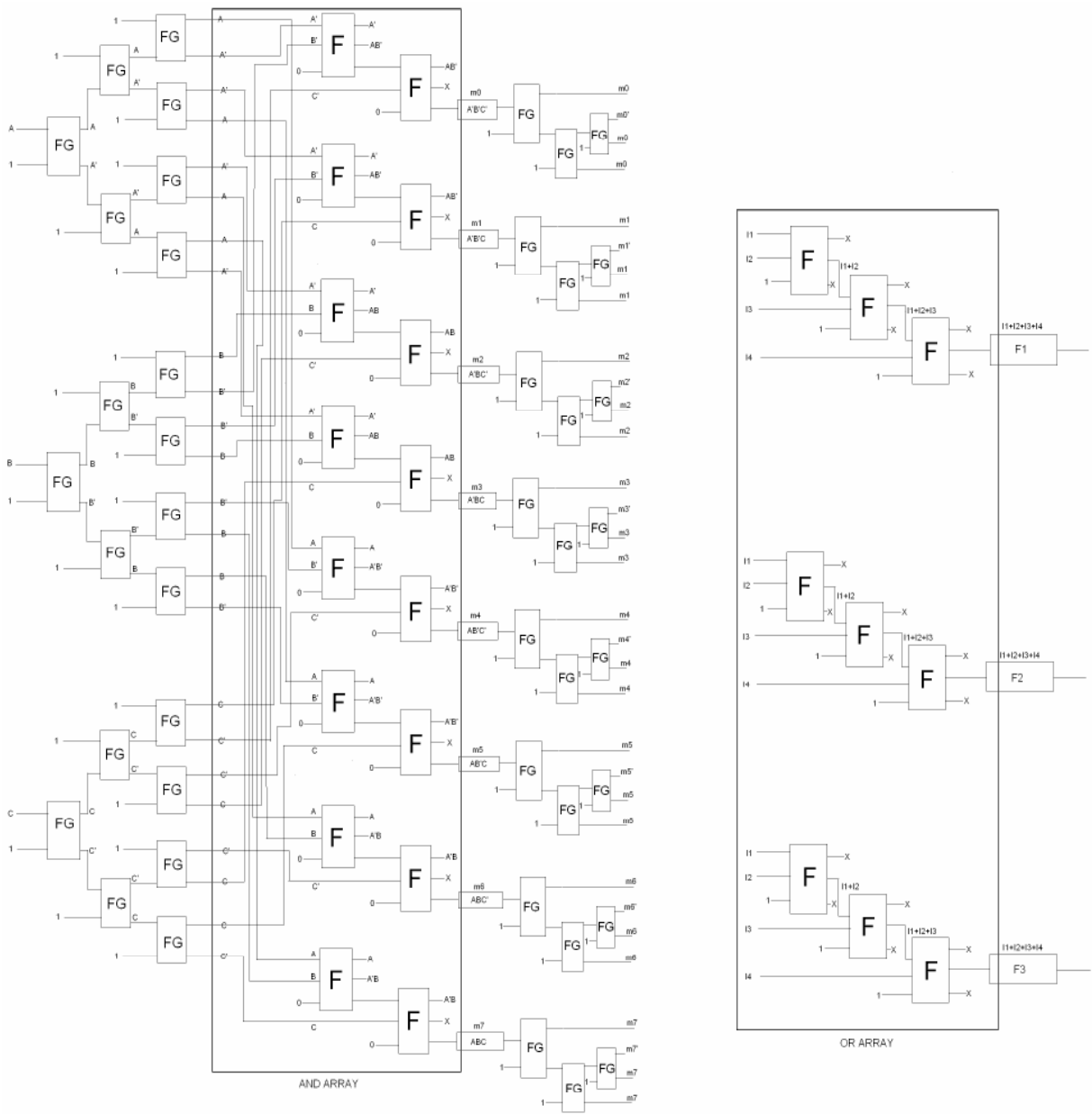

Figure 5. General 3 input RPLA Designed Using Fredkin(F) and Feynman(FG) Gates

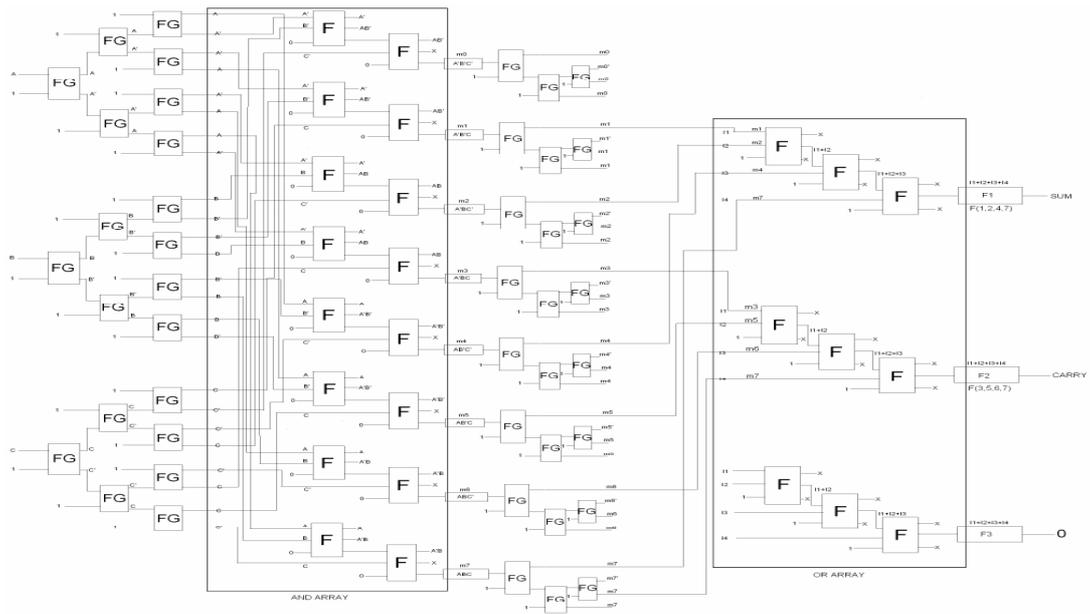

Figure 6.  Realization of Full Adder Using Designed 3 Input RPLA

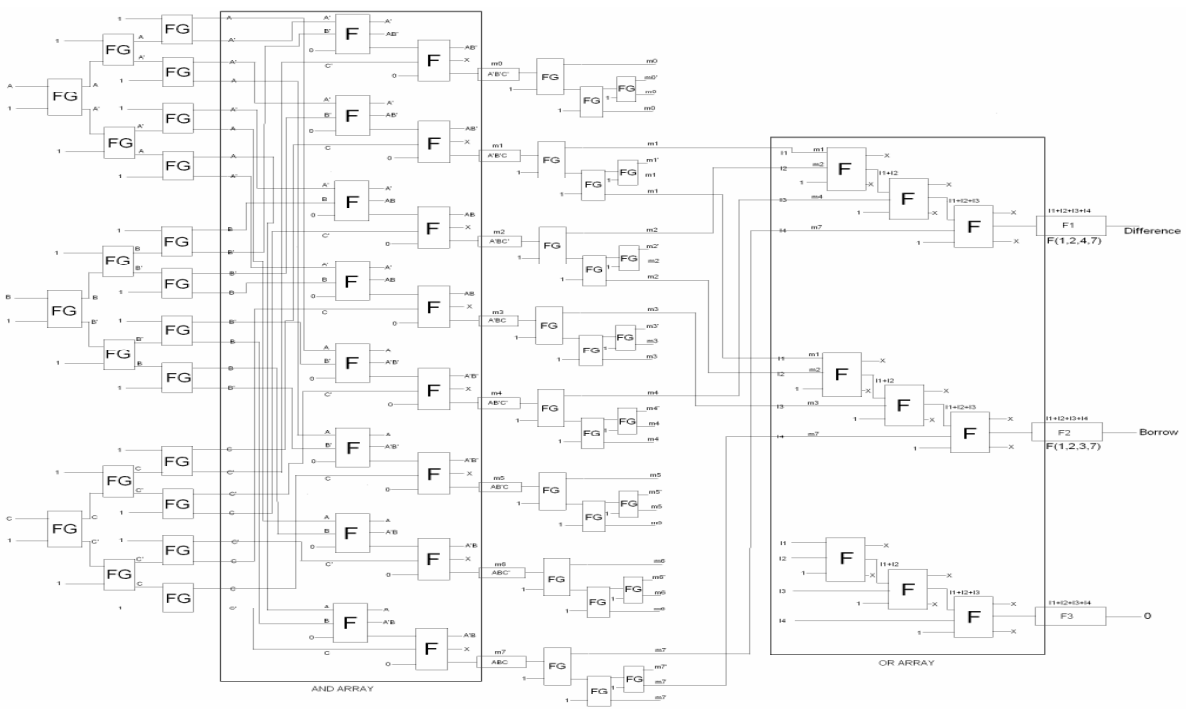

Figure 7.  Realization of Full Subtractor Using Designed 3 Input RPLA

*B. Evaluation of the Proposed RPLA*

The proposed reversible programmable logic array (RPLA) is highly optimized in terms of number of reversible gates and garbage outputs. As far as literature and our knowledge is concerned, among the existing reversible gates Fredkin gate is most suitable to generate the AND and OR functions, with minimum gates and garbage outputs. Furthermore, in the proposed RPLA complementing and copying of the signals are required. This can be performed through a number of existing reversible gates, but the authors have used the Feynman gate, since it complements and copies the signal with minimum garbage output. In the Feynman gate, there are exactly two outputs corresponding to the inputs and a '0' in the second input will copy the first input to both the outputs of that gate and a '1' in the second input will produce the complement of the input at the second output and pass the same input at the first output as shown in Figure 3. Thus, both the outputs can now be utilized resulting in zero garbage output. Hence it can be concluded that the proposed RPLA is highly optimized in terms of number of reversible gates and garbage outputs.

## V.   CONCLUSIONS

The focus of this paper is on the proposal of reversible programmable logic array (RPLA) using reversible gates. The 3 input RPLA which can generate any $2^8$ functions using the combination of 8 min terms ($2^3$) is also successfully designed. Finally, the application of the RPLA is demonstrated by implementing the reversible 1-bit full adder and subtractor. It is also demonstrated that the proposed design is highly optimized in terms of number of reversible gates and garbage outputs. It is expected that the proposed work will provide to a new paradigm to the arena of reconfigurable computing.